\documentclass[%
reprint,
superscriptaddress,
amsmath,amssymb,
aps,
pra,
]{revtex4-1}

\usepackage[utf8]{inputenc}
\usepackage{graphicx,import}   
\usepackage[colorlinks]{hyperref}
\usepackage{xcolor}
\usepackage{amsmath}
\usepackage{SIunits}
\usepackage{chngcntr}

\usepackage[colorinlistoftodos,prependcaption,textsize=tiny]{todonotes}
\usepackage{multirow}

\definecolor{lightgray}{RGB}{210,210,210}
\definecolor{bluegray}{RGB}{40,180,160}
\definecolor{navygray}{RGB}{110,140,170}
\definecolor{pp}{RGB}{140,0,211}
\definecolor{oo}{RGB}{255,165,0}
\definecolor{ff}{RGB}{255,0,255}
\definecolor{meadowgreen}{RGB}{0,128,0}
\definecolor{coolbrown}{RGB}{165,42,42}

\newcommand{\be}{\begin{equation}}
\newcommand{\ee}{\end{equation}}
\newcommand{\ba}{\begin{equation}}
\newcommand{\ea}{\end{equation}}
\newcommand{\bea}{\begin{eqnarray}}
\newcommand{\eea}{\end{eqnarray}}

\newcommand{\eref}[1]{Eq.~(\ref{#1})}

\DeclareUnicodeCharacter{2212}{-}
\hypersetup{
citecolor={bluegray}, %references
linkcolor={navygray} %equations
}

\begin{document}

\title{Interplay between kinetic inductance, non-linearity \\ and quasiparticle dynamics in granular aluminum MKIDs}

\author{Francesco~Valenti}
\affiliation{Physikalisches Institut, Karlsruhe Institute of Technology, 76131 Karlsruhe, Germany}

\affiliation{ \mbox { Institut f{\"u}r Prozessdatenverarbeitung und Elektronik,~Karlsruhe Institute of Technology}, 76344 Eggenstein-Leopoldshafen, Germany}

\author{F{\'a}bio~Henriques}
\affiliation{Physikalisches Institut, Karlsruhe Institute of Technology, 76131 Karlsruhe, Germany}

\author{Gianluigi~Catelani}
\affiliation{ \mbox{JARA Institute for Quantum Information (PGI-11),~Forschungszentrum J{\"u}lich, 52425 J{\"u}lich, Germany}}
%mbox needed to avoid linebreak

\author{Nataliya~Maleeva}
\affiliation{Physikalisches Institut, Karlsruhe Institute of Technology, 76131 Karlsruhe, Germany}

\author{Lukas~Gr\"unhaupt}
\affiliation{Physikalisches Institut, Karlsruhe Institute of Technology, 76131 Karlsruhe, Germany}

\author{Uwe~von~L\"upke}
\affiliation{Physikalisches Institut, Karlsruhe Institute of Technology, 76131 Karlsruhe, Germany}

\author{Sebastian~T.~Skacel}
\affiliation{Institute of Nanotechnology, Karlsruhe Institute of Technology, 76344 Eggenstein-Leopoldshafen, Germany}

\author{Patrick~Winkel}
\affiliation{Physikalisches Institut, Karlsruhe Institute of Technology, 76131 Karlsruhe, Germany}

\author{Alexander~Bilmes}
\affiliation{Physikalisches Institut, Karlsruhe Institute of Technology, 76131 Karlsruhe, Germany}

\author{Alexey~V.~Ustinov}
\affiliation{Physikalisches Institut, Karlsruhe Institute of Technology, 76131 Karlsruhe, Germany}
\affiliation{Russian Quantum Center, National University of Science and Technology MISIS, 119049 Moscow, Russia}

\author{ Johannes~Goupy }
\affiliation{Universit{\'e} Grenoble Alpes, CNRS, Grenoble INP, Institut N{\'e}el, 38000 Grenoble, France}

\author{ Martino~Calvo }
\affiliation{Universit{\'e} Grenoble Alpes, CNRS, Grenoble INP, Institut N{\'e}el, 38000 Grenoble, France}

\author{ Alain~Beno\^{i}t}
\affiliation{Universit{\'e} Grenoble Alpes, CNRS, Grenoble INP, Institut N{\'e}el, 38000 Grenoble, France}

\author{Florence~L{\'e}vy-Bertrand}
\affiliation{Universit{\'e} Grenoble Alpes, CNRS, Grenoble INP, Institut N{\'e}el, 38000 Grenoble, France}

\author{Alessandro~Monfardini}
\affiliation{Universit{\'e} Grenoble Alpes, CNRS, Grenoble INP, Institut N{\'e}el, 38000 Grenoble, France}

\author{Ioan~M.~Pop}
\email{ioan.pop@kit.edu}
\affiliation{Physikalisches Institut, Karlsruhe Institute of Technology, 76131 Karlsruhe, Germany}

\affiliation{Institute of Nanotechnology, Karlsruhe Institute of Technology, 76344 Eggenstein-Leopoldshafen, Germany}

\date{\today}

\begin{abstract}
Microwave kinetic inductance detectors (MKIDs) are thin film, cryogenic, superconducting resonators. Incident Cooper pair-breaking radiation increases their kinetic inductance, thereby measurably lowering their resonant frequency. For a given resonant frequency, the highest MKID responsivity is obtained by maximizing the kinetic inductance fraction $\alpha$. However, in circuits with $\alpha$ close to unity, the low supercurrent density reduces the maximum number of readout photons before bifurcation due to self-Kerr non-linearity, therefore setting a bound for the maximum $\alpha$ before the noise equivalent power (NEP) starts to increase. By fabricating granular aluminum MKIDs with different resistivities, we effectively sweep their kinetic inductance from tens to several hundreds of pH per square. We find a NEP minimum in the range of $25\; \text{aW}/\sqrt{\text{Hz}}$ at $\alpha \approx  0.9$, which results from a trade-off between the onset of non-linearity and a non-monotonic dependence of the noise spectral density vs. resistivity.
\end{abstract}

\maketitle

\section{Introduction}

Since their first implementation fifteen years ago \cite{day2003broadband}, microwave kinetic inductance detectors (MKIDs) play an important role in ground based radioastronomy \cite{monfardini2014latest, schlaerth2008millimeter, maloney2010music, shirokoff2012mkid,mazin2013arcons, oguri2016groundbird}, particle detection \cite{quaranta2013x, szypryt2015ultraviolet,battistelli2015calder, cardani2015energy,cardani2017high}, and are promising candidates for spaceborne millimeter wave observations \cite{patel2013fabrication, matsumura2014mission, griffin2016spacekids,baselmans2016performance}. This remarkable development was facilitated by the technological simplicity of MKIDs, which consist of thin film, cryogenic, superconducting resonators in the microwave domain. The MKID signal is the shift of its resonant frequency due to an increase in the kinetic inductance of the film, which is itself proportional to the number of Cooper pairs (CP) broken by the incoming radiation. These compact, low-loss and multiplexable detectors can also provide a convenient tool to probe material properties, such as the change of dielectric constant due to a superfluid transition \cite{grabovskij2008situ}, the density of states of granular aluminum and indium oxide \cite{dupre2017tunable}, or to image phonons \cite{swenson2010high,moore2012position}.

\begin{figure}[!t]
\begin{center}
\vspace{1 mm}
\def\svgwidth{1 \columnwidth}  
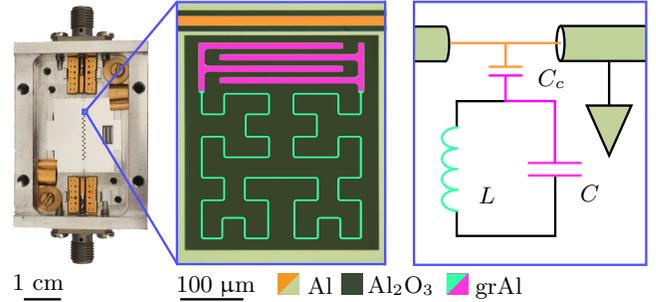
\caption{Optical images of the aluminum sample holder and a grAl lumped element resonator, together with its equivalent circuit. A sample consists of an ensemble of 22 resonators coupled to a central coplanar microwave waveguide, which is used to perform transmission spectroscopy. The zoom-in shows a single resonator, where we highlight the interdigitated capacitor in magenta (which gives both $C$ and $C_c$) and the meandered inductor (which gives $L$) in green. The sapphire substrate is shown in dark green. All resonators are fabricated from a $20$ nm thick grAl film using e-beam lift-off lithography, while the central conductor of the coplanar waveguide (orange) and the ground plane (light green) are made of $50$ nm thick aluminum patterned by optical  lift-off. The meandered inductor for samples A-D is shaped as a third degree Hilbert curve (shown), while for samples E and F it is shaped as a second degree Hilbert curve (cf. text and Appendix~\ref{hilbertapp}). In order to distribute the resonant frequencies $f_0 = (2\pi \sqrt{LC})^{-1} $ of the 22 resonators, we sweep the capacitances by changing the length of the interdigitated fingers.}
\label{fig:sample}
\end{center}
\end{figure}

The first MKIDs consisted of thin film aluminum distributed element resonators. Their numerous incarnations now include lumped element resonator geometries \cite{doyle2008lumped}, novel solutions such as spiral resonators \cite{hayashi2013microwave} or various kinds of antenna coupling \cite{naruse2013optical,sekimoto2016design,sayers2010optics}, and a wealth of different film materials such as TiN \cite{gao2012titanium,swenson2013operation}, NbN \cite{saito2015relationship}, PtSi \cite{szypryt2016high} and W$_x$Si$_y$ \cite{cecil2012tungsten}, including hybrid realizations \cite{janssen2014performance} and  multilayered films \cite{catalano2015bi,dominjon2016study,cardani2018ti}.

\newpage 

Here, we propose granular aluminum (grAl), a composite material made of pure Al grains with median size of the order of $a = 3\pm 1$~nm in an aluminum oxide matrix \cite{abeles1966enhancement,deutscher1973granular}, as a novel material for MKIDs. As illustrated in Fig. \ref{fig:sample}, we employ a coplanar waveguide (CPW) geometry, in which the ground plane and the feedline of the CPW are made of pure aluminum, $50$ nm thick, and the lumped element MKID resonators are entirely made of grAl.

Granular aluminum is an appealing material because it already demonstrated high internal quality factors - in the range of $10^6$ - in the microwave domain \cite{sun2012measurements,rotzinger2016aluminium,grunhaupt2018dynamics}, and ease of fabrication, which simply consists in aluminum deposition in a controlled oxygen atmosphere \cite{deutscher1973granular,abeles1966enhancement}. Furthermore, by varying the oxygen pressure during deposition, one can tune material parameters such as the resistivity (from $1$ to $10^4 \; \upmu \Omega \cdot$cm), kinetic inductance, and superconducting gap. The kinetic inductance of a square of thin film is determined by the ratio of the normal state sheet resisitance $R_n$ and the  critical temperature $T_c$ \cite{tinkham1996introduction,annunziata2010tunable}:
\begin{equation}\label{lkinSquare}
L_{\Box} = \frac{0.18 \hbar}{k_B} \frac{R_n}{T_c},
\end{equation}
and, in the case of grAl, it can reach values as high as a few $\text{nH/}\Box$ \cite{rotzinger2016aluminium,grunhaupt2018dynamics}.

The grAl microstructure, consisting of superconducting grains separated by thin insulating shells, can be modeled as a network of Josephson junctions (JJ),  which simplifies to a 1D JJ chain for resonators in the thin ribbon limit (length~$\gg$~width~$\gg$~thickness) \cite{maleeva2018circuit}. We can use this model to quantitatively estimate the non-linearity of grAl resonators and extract the self-Kerr coefficient of the fundamental mode $K_{11}$ \cite{maleeva2018circuit}, similarly to the case of resonators made of mesoscopic JJ arrays \cite{bourassa2012josephson,tancredi2013bifurcation,weissl2015kerr}.

This article is organized as follows: In Section~\ref{sec:theory}, we propose a model to quantify the interplay between kinetic inductance fraction, non-linearity and quasiparticle dynamics, and its effect on the noise equivalent power (NEP) - the main operational figure of merit of MKIDs. We discuss experimental methods and results in Section~\ref{sec_exp}, showing that we can exploit the tunability of the grAl non-linearity and superconducting gap to achieve low NEP values. In Section~\ref{sec:conclusions} we conclude by proposing guidelines to further reduce the NEP in future designs.

\section{Theory} \label{sec:theory}

The NEP is defined as the radiant power needed to have equal signal and noise amplitudes in a $1$ Hz output bandwidth, and can be expressed as \cite{catalano2015bi}
\begin{equation}\label{nepeq}
\text{NEP} = \frac{\text{NSD}}{\mathfrak{R}},
\end{equation} 
where NSD is the noise spectral density and $\mathfrak{R}$ is the responsivity. In the following subsections we will discuss in detail the influence of grAl parameters on each of these two quantities.

\subsection{Noise spectral density}
Fluctuations of the MKID resonant frequency in the absence of incoming radiation constitute noise. The NSD is computed by recording the fluctuations of the resonant frequency over time, taking its Fourier transform and dividing by the square root of the output bandwidth, hence the NSD is quoted in Hz$/\sqrt{\text{Hz}}$ units (cf. Appendix~\ref{NSDapp}).

These fluctuations can be either dominated by the added noise of the measurement setup, or by noise sources intrinsic to the resonator, such as dielectric two level systems \cite{muller2017towards}, microscopic charge fluctuators \cite{sueur2018microscopic}, or fluctuations in the number of quasiparticles (QPs) \cite{mazin2005microwave,sergeev1996photoresponse,sergeev2002ultrasensitive}. As discussed in Section~\ref{SecNEPmeasurement}, for the grAl MKIDs in this work the latter mechanism appears to be dominant, and reducing the QP density can be a fruitful approach to suppress the NSD. 

\subsection{Phonon trapping and quasiparticle fluctuations} \label{sec2PhononTrapping}

\begin{figure}[t!]
\begin{center}
\hspace{-0.8 cm} 
\def\svgwidth{1 \columnwidth}  
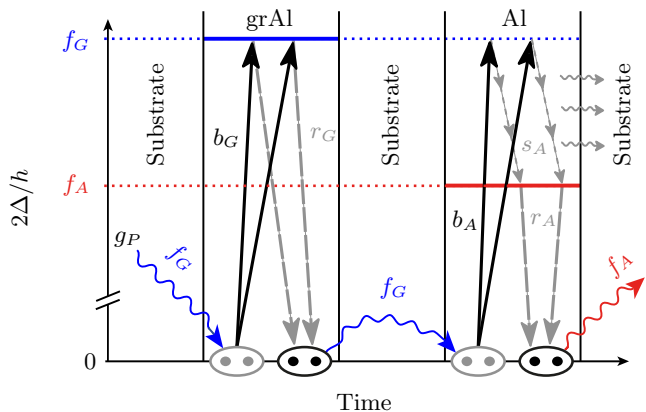
\caption{Phonon trapping and quasiparticle number reduction: schematic depiction of the dynamics described in Eqs.~\eqref{dotng}~to~\eqref{dotnp}. We show the superconducting energy band diagram in the excitation picture for both grAl (in blue) and Al (in red). Excitations and relaxations are represented in solid black and dashed gray arrows respectively. Wiggly arrows represent phonons and the labels represent their corresponding frequency. Substrate phonons with characteristic frequency larger than the grAl spectral gap $f_G$, generated at rate $g_P$, can break grAl CPs at rate $b_G$. The resulting grAl QPs recombine at rate $r_G$, emitting phonons that can travel through the substrate and reach the Al ground plane, which covers most of the chip (cf. Fig.~\ref{fig:sample}), breaking Al CPs at rate $b_A$. The excited QPs can scatter to a lower energy $h f_A$ via electron-phonon interaction at rate $s_A$ and recombine to form an Al CP at rate $r_A$. In both cases the emitted phonons have characteristic frequencies lower that the spectral gap of grAl, thus being unable to break grAl CPs.}
\label{fig:trapping}
\end{center}
\end{figure}

The grAl MKIDs are surrounded by the comparatively much larger aluminum ground plane (cf. Fig. \ref{fig:sample}), which has a lower superconducting gap and can act as a phonon trap  \cite{monfardini2016lumped,baselmans2017kilo}, possibly reducing the number of QPs generated by nonthermal phonons from the substrate, as schematically illustrated in Fig \ref{fig:trapping}.

In order to model the effect of phonon traps, we start from the expression of the NEP (in the ideal case of unit conversion efficiency) dominated by QP generation-recombination noise \cite{sergeev1996photoresponse,sergeev2002ultrasensitive},
\begin{equation} \label{grnepeq}
\text{NEP} = 2 \Delta_G \sqrt{\frac{N_G}{\tau_G}},
\end{equation}
where $\Delta_G$, $N_G$ and $\tau_G$ are the grAl superconducting gap, QP number, and QP lifetime respectively. In the following we use the indexes $G$ and $A$ to refer to thin film grAl and aluminum. We assume that all phonons in grAl, Al and the substrate quickly reach the steady state after a high energy generation event \cite{swenson2010high,moore2012position}, allowing us to described their density as position independent. Since the temperature $T$ satisfies $1.76 k_B T \ll \Delta_G, \Delta_A, \Delta_G-\Delta_A$ we focus only on the ``{hot}" phonons with energy $h f_P \geq \Delta_{G}$. We define the number of phonons in the substrate $N_P$ and write a dynamical system of the Rothwarf-Taylor type \cite{rothwarf1967measurement} 
\begin{align}
&\dot{N}_G = -2 r_G N_G^2 + 2b_GN_P \label{dotng}, \\
&\dot{N}_A = -2r_AN_A^2 + 2b_AN_P-s_AN_A  \label{dotna},  \\
&\dot{N}_P = g_P - b_AN_P + r_AN_A^2 - b_GN_P + r_GN_G^2 \label{dotnp},  
\end{align}
where $r$ and $s$ are the rates of QP recombination and scattering respectively, $b$ is the rate of CP breaking, and $g$ is a phonon generation rate due to external processes (e.g. high energy impacts). Solving the system in the steady state under the assumption of weak scattering $s_A \ll 2 \sqrt{r_A g_P}$ and plugging the result into Eq.~\eqref{grnepeq} we can write the fractional change in NEP as a function of the fractional change in $\Delta_G$ as 
\begin{equation} \label{neppr}
\frac{\delta \text{NEP}}{ \overline {\text{NEP}}} = - \left(  \frac{3\overline{\Delta_G}+4\Delta_A}{2(\overline{\Delta_G}-\Delta_A)}  \right) \frac{\delta \Delta _G}{  \overline{\Delta_G}   },\end{equation}
where overline bars denote average values. The reader is invited to refer to Appendix~\ref{trappingmodel} for a more detailed description of the model. Equation~\eqref{neppr} indicates an anticorrelation between the NEP and the superconducting gap $\Delta_G$. In Section \ref{SecNEPmeasurement} we present experimental evidence for this anticorrelation, and we show that the increase of the grAl gap at the top of the so-called superconducting dome \cite{abeles1966enhancement,deutscher1973granular} is responsible for a significant improvement in detector performance. 
\subsection{Responsivity}
The responsivity of a MKID can be expressed as \cite{day2003broadband,catalano2015bi, wang2014measurement,grunhaupt2018dynamics}
\begin{equation} \label{respeq}
\mathfrak{R} = \frac{|\delta f_0|}{P_\text{abs}} = \alpha f_0 \frac{\delta x_\text{qp}}{P_\text{abs}},
\end{equation}
where $\delta f_0$ is the shift in resonant frequency due to pair breaking,  $P_\text{abs}$ is the radiant power absorbed by the detector,  $\alpha$ is the kinetic inductance fraction with respect to the total inductance, $f_0$ is the unperturbed resonant frequency, and $\delta x_\text{qp}$ is the shift in the quasiparticle density, defined as twice the fraction of broken CPs. For practical reasons, the choice of values for the operational frequencies $f_0$ is limited by the availability and cost of readout electronics, and it is typically in the range of a few GHz. To estimate $P_\text{abs}$, we assume that every collected photon with $h f > 2 \Delta $ breaks a CP. The amount of impedance matching between the resonator plane and the medium through which the photons propagate (for example vacuum or dielectric substrate) defines the detector absorptance $\mathcal{A}$ (cf. Appendix~\ref{hilbertapp}). The power absorbed by the resonators is then $P_\text{abs}=\delta P_\text{in}\cdot \mathcal{A}$, where $\delta P_\text{in}$ is the change in radiant power under illumination through the optical setup. Under operational conditions, for any MKID design, one aims to minimize the impedance mismatch for the incoming radiation, in order to obtain a value for $\mathcal{A}$ as close as possible to unity.

\subsection{Voltage responsivity}

\begin{figure*}%[!tb]
\begin{center}
\def\svgwidth{\textwidth}  
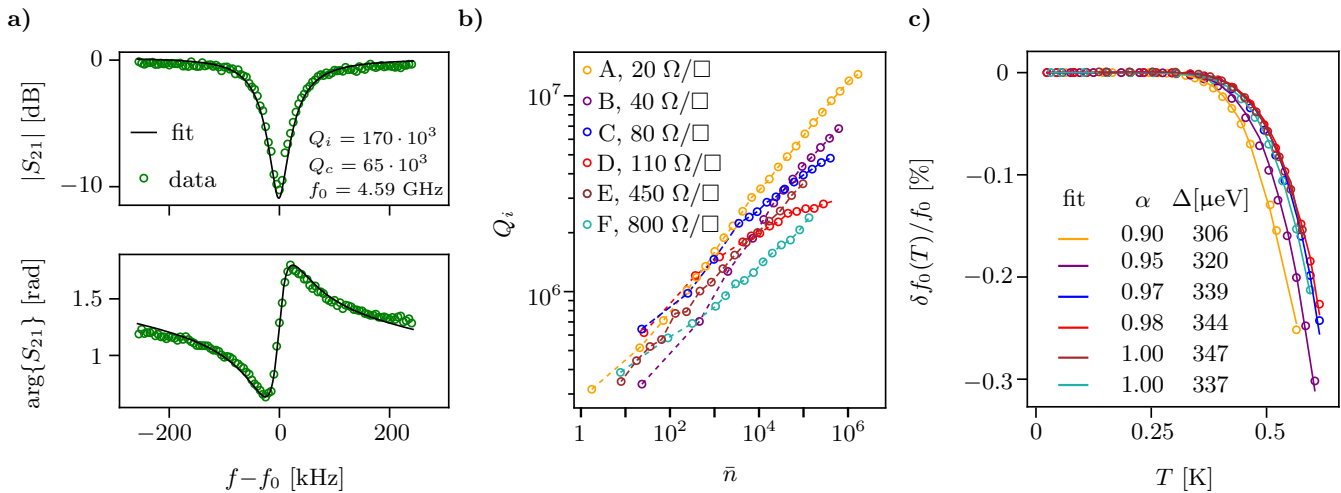
\caption{Microwave characterization of MKIDs via transmission measurements. 
\textbf{a)} Amplitude (normalized by the sample holder response) and phase of the transmission coefficient $S_{21}$ for a resonator in sample A, at a readout power in the single photon regime ($\bar{n}\approx 1$).  Raw data is shown as green circles and the solid black line is the fit to the complex scattering parameters from Ref.~\cite{probst2015efficient}; the fitted values are given in the top panel.
\textbf{b)} Internal quality factors as a function of the average number of readout photons circulating in the resonator, shown up to the bifurcation threshold $\bar{n} = n_\text{max}$ in log-log scale for all samples. The average photon number is calculated from the estimated on-chip readout power as $\bar{n} = 2 Q_c P_\text{cold} / \hbar \omega_0^2$ (cf. Appendix~\ref{nmaxapp}). The single photon regime corresponds to $P_\text{cold} \approx -150$ dBm. Notice that the maximum photon number decreases for higher resisitivity films, due to the higher non-linearity (cf. Eq.~\eqref{nmaxeq}).
\textbf{c)} Change in resonant frequency as a function of temperature, averaged for all resonators in each sample (colored circles, using the same palette as in panel b) ). We fit the measured values with the BCS equation ${\delta f_0 (T)} / {f_0} = -{\alpha} \sqrt{ 2{\pi \Delta}/{k_B T} }\exp \left( {-{\Delta}/{k_B T}} \right)$ from Ref.~\cite{gao2008physics} (colored lines), where we use $f_0 = f_0(T\approx 25 \; \text{mK})$. The fitting parameters are the kinetic inductance fraction $\alpha$ and the superconducting gap $\Delta$, shown in the legend.}
\label{fig:mwcharacteriz}
\end{center}
\end{figure*}

Following Eq. \eqref{respeq}, the responsivity $\mathfrak{R}$ scales linearly with the kinetic inductance fraction $\alpha$. However, in the limit $\alpha \to 1$ the performance of MKIDS is limited by the early onset of non-linear phenomena, i.e. the resonators bifurcate at low readout voltages. One wants to operate MKIDs at the highest possible readout power before bifurcation \cite{duffing1918forced,eichler2014controlling,maleeva2018circuit}, in the range of $10^5-10^6$ circulating photons (cf. Fig.~\ref{fig:mwcharacteriz}), in order to maximize the microwave signal to noise ratio. The NEP defined in Eq.~\eqref{nepeq} is thus \textit{implicitly} dependent on the maximum microwave readout voltage, which in turn scales with the square root of the maximum number of circulating photons before bifurcation $n_\text{max}$. This dependence can be made explicit by defining a \textit{voltage responsivity}
\begin{equation} \label{voltresp}
\mathfrak{R}_V \equiv \alpha \sqrt{n_\text{max}}.
\end{equation}
The voltage responsivity is trivially zero if $\alpha$ is zero, but it also vanishes in the limit $L_\text{kin} \to \infty$ ($\alpha \to 1$), when the resonator non-linearity also increases, suppressing $n_\text{max}$. In the following, we quantify this non-monotonic dependence. 

\begin{figure*}%[!tb]
\begin{center}
\def\svgwidth{\textwidth}  
\hspace{-1cm}
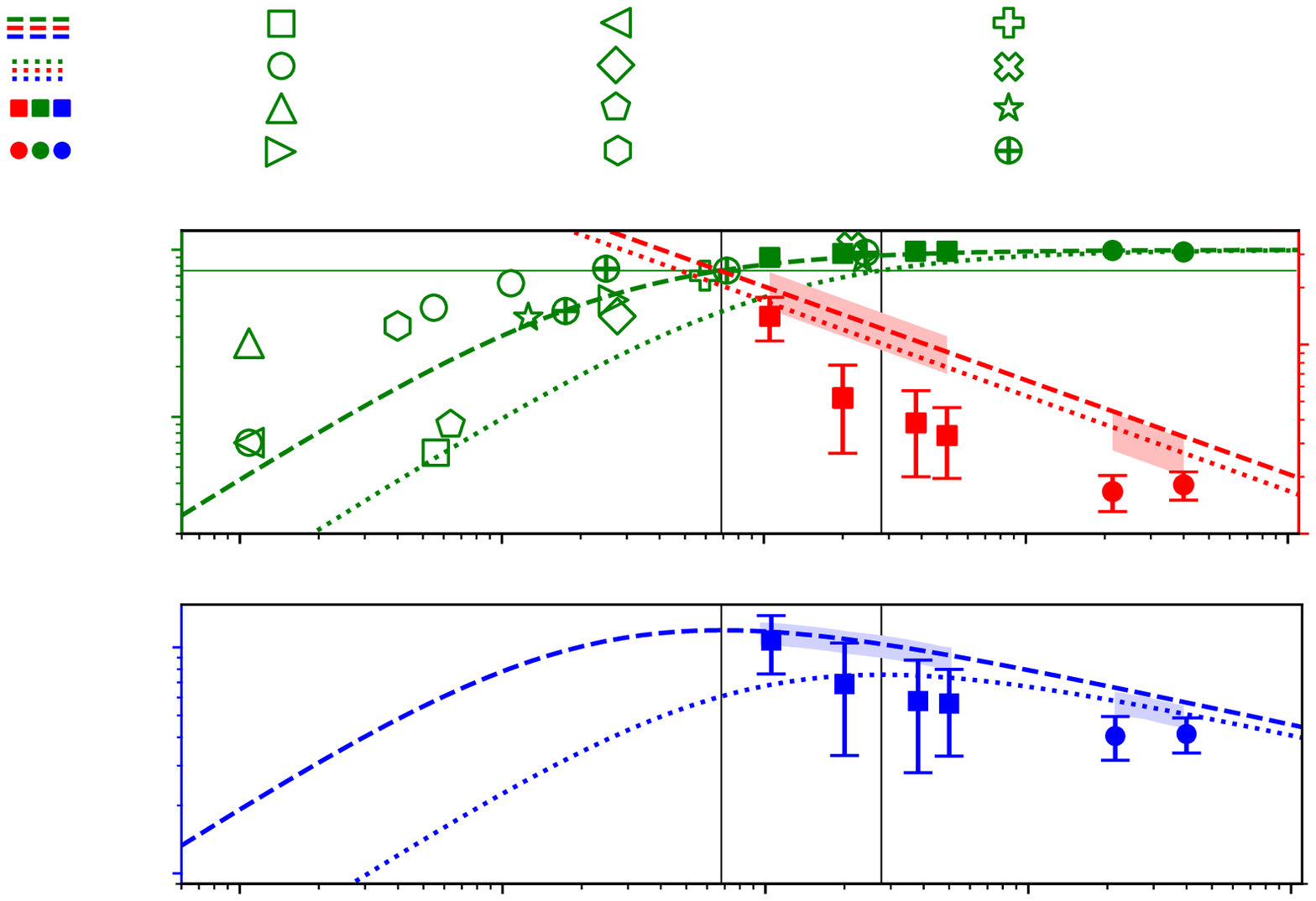
\caption{Kinetic inductance fraction, maximum number of readout photons and voltage responsivity as a function of the sheet resistance per square $R_n$ scaled to the critical temperature $T_c$. \textbf{a)} We plot the kinetic inductance fraction $\alpha$ in green and the maximum number of readout photons before bifurcation $n_\text{{max}}$ in red. Dashed and dotted lines are analytical predictions for H3 and H2 geometries used in this work, for both $\alpha$ (cf. Appendix~\ref{alphaapp}) and $n_\text{max}$ (estimated with Eq.~\eqref{nmaxeq} at a fixed $Q_c = 10^5$). The color shaded regions overlapping with the $n_\text{max}$ analytical lines represent the range of maximum photon number values corresponding to the different interdigitated capacitances $C$ of the various resonators in each sample (cf. Eq.~\eqref{nmaxeq}). Full markers show values measured in this work: squares and circles relate to H3 and H2 geometries, respectively. For comparison, empty markers show $\alpha$ values reported in the literature for various materials (cf. Appendix \ref{alphaapp}). The measured $n_\text{max}$ values are approximately two times lower than predicted for all samples, which might be due to a systematic underestimation of the on-sample readout power. The reported values are averaged over all functional resonators for each sample. \textbf{b)} Voltage responsivity $\mathfrak{R}_V$, defined as the product between the kinetic inductance fraction $\alpha$ and the square root of the maximum number of readout photons before bifurcation $n_\text{max}$ (cf. Eq.~\eqref{voltresp}). We report values for H3 and H2 geometries with dashed/dotted lines (analytical predictions) and square/circle markers (measured data) respectively. The color shaded regions overlapping with the lines represent the range of different interdigitated capacitances $C$. The maximum voltage responsivity is obtained at $\alpha = 3/4$ (cf. panel~a)), as indicated by the vertical black lines across the two panels (left line for H3 geometry, right line for H2).}
\label{fig:alphanvoltresp}
\end{center}
\vspace{- .5 cm}
\end{figure*}

The kinetic inductance fraction $\alpha = L_\text{kin} / L_\text{tot}$ can be estimated by knowing the geometry, resistivity and critical temperature of a resonator, as we discuss in Appendix~\ref{alphaapp}. We show that for a grAl lumped element resonator with kinetic inductance dominating over the geometric inductance we can write (cf. Appendix~\ref{nmaxapp})%
\begin{equation} \label{nmaxeq}
n_\text{max} = \frac{4 \ell^2 \hbar \sqrt{C}}{ 3\sqrt{3}Q_c (\pi e a)^2 \sqrt{L_\text{kin}}},
\end{equation}
where $\ell$ is the meandered inductor length and $C$ is the interdigitated capacitance (cf. Fig.\ref{fig:sample}), $Q_c$ is the coupling quality factor, $e$ is the electron charge, and $a$ is the size of an aluminum grain. By plugging $\alpha$ and Eq.~\eqref{nmaxeq} into Eq.~\eqref{voltresp} one finds \begin{equation} \label{rv2}
\mathfrak{R}_V \sim \frac{L_\text{kin}^{3/4}}{L_\text{kin}+L_\text{geom}},
\end{equation}with $L_\text{kin} \sim R_n/T_c$ (cf. Eq.  \eqref{lkinSquare}), where $R_n$ is the normal state sheet resistance per square, and $T_c$ is the critical temperature. From Eq.~\eqref{rv2} one can see that $\mathfrak{R}_V$ tends to zero for both limit cases $L_\text{kin} = 0$ (no detection) and $L_\text{kin} \to \infty$ (fully non-linear system). The voltage responsivity increases sharply with $L_\text{kin}$ until it reaches a maximum at $\alpha=3/4$, after which it slowly decreases (cf. Fig.~\ref{fig:alphanvoltresp}). Thus, $\mathfrak{R}_V$ quantifies the interplay between kinetic inductance and non-linearity, and is a convenient metric to determine whether a given kinetic inductance fraction $\alpha$ gives a sensitive enough detector, while not being severely limited by non-linearity.

\section{Experimental results} \label{sec_exp} 

\subsection{Measurement setups}

The MKIDs discussed in this work are lumped element resonators, composed of a meandered inductor shaped as a Hilbert curve of third or second degree (H3 and H2), and an interdigitated capacitor (cf. Appendix~\ref{hilbertapp}). The $20$~nm thick grAl film is patterned on a sapphire substrate via e-beam lift-off lithography. We fabricate $2\times2$~cm$^2$ chips hosting $22$ resonators each. We label the chips from A to F according to the grAl sheet resistance, of 20, 40, 80, 110, 450, 800 $\Omega / \Box$, respectively. The 50 nm thick surrounding Al CPW ground plane is patterned in a second optical lift-off lithography step.

Two different measurement setups are required in order to characterize a) the intrinsic properties of the MKID resonators, at frequencies in the GHz range, and b) the operational MKID response to millimeter wave radiation. To measure their microwave properties, we anchor the MKIDs to the base plate of a so-called ``{dark}" dilution cryostat, with a base temperature of about $25$~mK. In this setup, we couple to room temperature electronics via heavily attenuated and filtered radio-frequency (RF) lines, including infrared (IR) filters (cf. Ref. \cite{grunhaupt2017argon}), with the goal of reducing stray radiation from the higher temperature stages of the cryostat. 

On the other hand, in order to measure the NEP, we need to shine millimeter wave radiation onto the resonators and operate them at high readout powers. We thus use a much less shielded dilution refrigerator, with an optical opening and a base temperature of approximately $150$~mK, which we refer to as the ``{optical}" cryostat. The sample chip is mounted in an aluminum box with an optical window on one side (cf. Fig. \ref{fig:sample} and Appendix \ref{sampleholderapp}). For measurements performed in the optical cryostat, the optical window is facing the mm-wave cryogenic optical setup \cite{catalano2015bi}. For measurements performed in the dark cryostat, the optical window is covered with aluminum tape, and the sample holder is placed in a series of  successive cryogenic infrared and magnetic shields, similar to Ref. \cite{grunhaupt2017argon}.

\subsection{Microwave characterization}

Figure \ref{fig:mwcharacteriz} shows the results of measurements performed in the dark cryostat. In Fig. \ref{fig:mwcharacteriz}a we show a typical result for the transmission coefficient $S_{21}$ 
in the vicinity of the resonant frequency of one of the MKIDs in sample A, for a readout power in the single photon regime. We employ the circle fit routine detailed in Ref.~\cite{probst2015efficient} to extract resonator parameters of interest, namely the resonant frequency $f_0$, the internal quality factor $Q_i$, and the coupling quality factor $Q_c$. An example of fitted on-resonance frequency response is shown in Fig.~\ref{fig:mwcharacteriz}a, from which we extract an internal quality factor on the order of $10^5$ in the single photon regime.

\begin{figure}[!htb]
\centering
\vspace{.5 cm}
\hspace{-0. cm} 
\def\svgwidth{1 \columnwidth}  
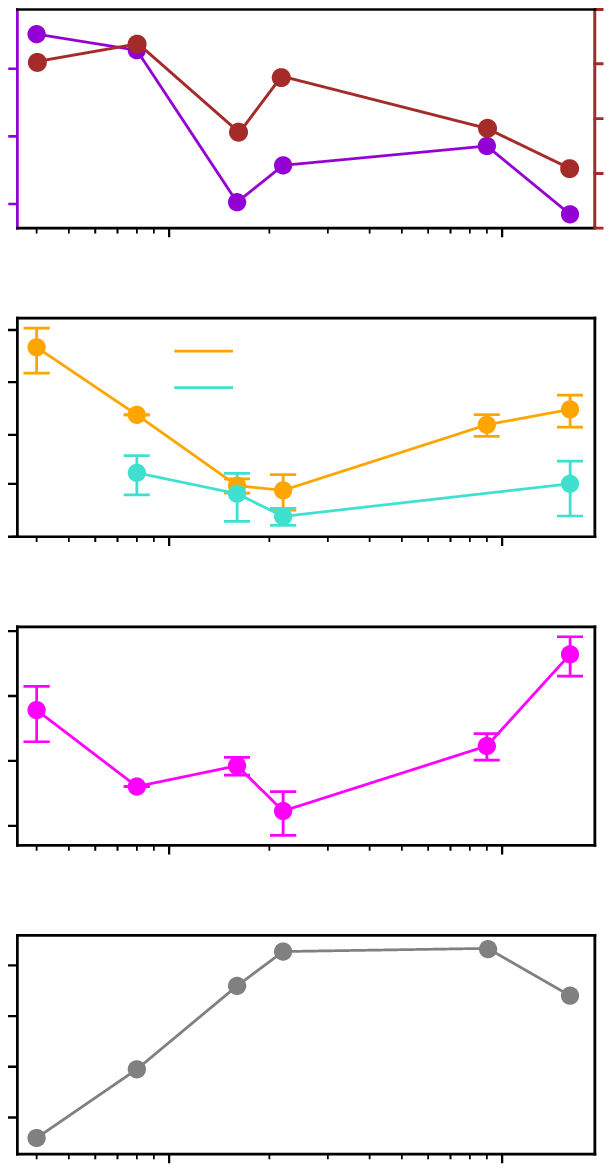
\caption{Properties of grAl MKIDs as a function of film resistivity. \textbf{a)} We use the measured values of $\delta f_0$, $f_0$, $\alpha$, and the estimated power absorbed by the resonators $P_\text{abs}$, to compute the responsivity and the shift in quasiparticle density  (cf. Eq.~\eqref{respeq}), which we plot in purple and brown respectively. \textbf{b)} Noise spectral density (NSD) measured for the same samples in what we denote ``optical" (in orange) and ``dark" (in turquoise) cryostats (cf. Section~\ref{sec_exp}), evaluated at $10$ Hz. The NSD shows a minimum at $\rho \approx 200\; \upmu\Omega\cdot$cm for measurements taken in both cryostats (see the main text and Appendix~\ref{NSDapp} for a detailed discussion). \textbf{c)} Noise equivalent power (NEP), calculated as the ratio of measured responsivity and NSD in the optical cryostat. The resistivity dependence of the NEP is dominated by that of the NSD. \textbf{d)} Measured grAl superconducting gap $\Delta$, extracted using the fitting procedure of Fig.~\ref{fig:mwcharacteriz}c. Note that values of NEP and $\Delta$ are anticorrelated (see text for a detailed discussion).}
\label{fig:detector_params}
\end{figure}

\begin{figure}[!htb]
\begin{center}
\def\svgwidth{1 \columnwidth}  
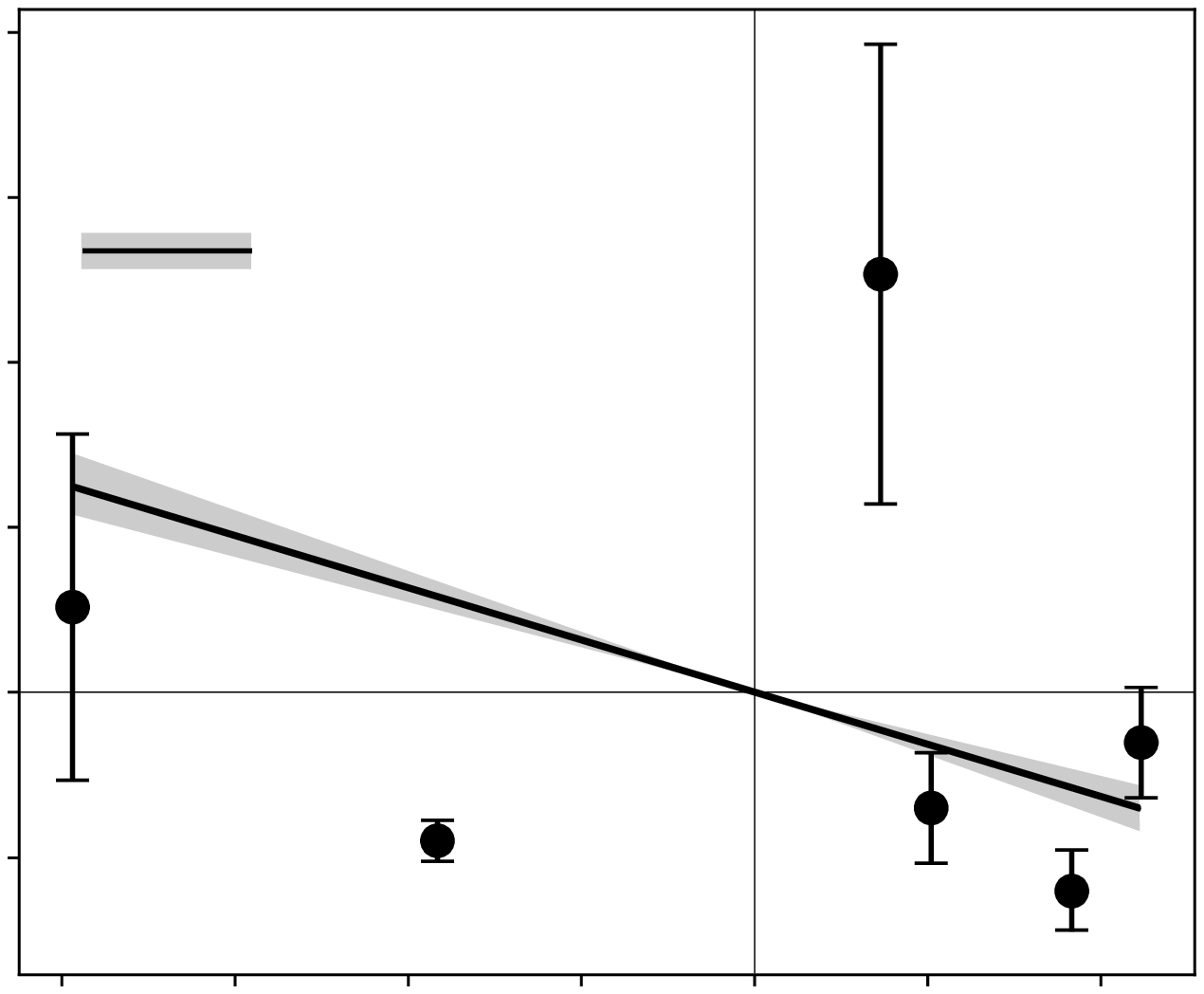
\caption{Fractional change in NEP as a function of the fractional change in $\Delta$. We represent the measured values with circular markers and the analytical prediction from Eq.~\eqref{neppr} as a solid black line, with a $10\%$ confidence interval shown as a shaded gray region.}
\label{FractionalChangeNEP}
\end{center}
\end{figure}

In order to compute the average number of readout photons in the resonator, $\bar{n}$, we estimate the on-chip power by summing the total attenuation on the input line of the cryostat (see Appendix~\ref{nbarapp}). It's important to note that, due to the uncertainity in the attenuation figure of the RF components over a broad frequyency range, this method is only accurate within an order of magnitude. In Fig. \ref{fig:mwcharacteriz}b we present measurements of $Q_i$ as a function of $\bar{n}$, for readout powers ranging from $\bar{n} \approx 1$ up to the critical number of readout photons $n_\text{max}$ (reported in Fig.~\ref{fig:alphanvoltresp}a along with its analytical calculation), at which the resonator bifurcates. The internal quality factor increases monotonically with the average number of photons in the resonators. This type of dependence was previously observed for grAl resonators in Ref.~\cite{grunhaupt2018dynamics}, and it is also routinely measured in thin film aluminum resonators \cite{muller2017towards}. The increase of the internal quality factor with power can be interpreted as the combined effect (cf. Fig.~2c from Ref. \cite{grunhaupt2018dynamics}) of dielectric losses saturation \cite{muller2017towards,zhang2018microresonators}, together with photon assisted unpinning of non-equilibrium QPs, which are then allowed to either diffuse away from regions of high current density or to recombine into CPs \cite{nsanzineza2014trapping,gustavsson2016suppressing}.

By sweeping the sample temperature up to about $600$~mK we observe a downward shift in the resonant frequency, as shown in Fig.~\ref{fig:mwcharacteriz}c, which we fit using a Bardeen-Cooper-Schrieffer model  \cite{gao2008physics} to obtain the kinetic inductance fraction $\alpha$ and the superconducting gap $\Delta$. In the inset of Fig.~\ref{fig:mwcharacteriz}c, we report the resulting fit parameters averaged over all functional resonators in each chip. The obtained $\alpha$ is in good agreement with the analytical prediction (cf. Fig.~\ref{fig:alphanvoltresp}a), and $\Delta$ shows a dome-shaped dependence with resistivity (cf. Figs.~\ref{fig:mwcharacteriz}c and~\ref{fig:detector_params}d) in agreement with Refs. \cite{abeles1966enhancement, cohen1968superconductivity,deutscher1973granular}. 

Using the model of Eq.~\eqref{nmaxeq} for the maximum number of circulating photons before bifurcation, in Fig.~\ref{fig:alphanvoltresp}a we plot the calculated $n_\text{max}$ and $\alpha$ vs. the ratio of the normal state sheet resistance and critical temperature, ${R_n}/{T_c}$, which determines the kinetic inductance (see Eq.~\eqref{lkinSquare}). For comparison, in Fig.~\ref{fig:alphanvoltresp}a we also overlay the measured values of $\alpha$ and $n_\text{max}$ for samples A-F, together with a literature survey of reported $\alpha$ values for superconducting resonators fabricated with various other materials.

By replacing the expressions of $\alpha$ and $n_\text{max}$ in Eq.~\eqref{voltresp}, we can compare analytical predictions of the voltage responsivity $\mathfrak{R}_V$ with measured values, as shown in Fig.~\ref{fig:alphanvoltresp}b. Considering that $T_c$ for grAl is almost constant, bounded to the $1.5-2.1$~K interval, for grAl resonators the voltage responsivity is almost exclusively a function of resistivity. Although $\mathfrak{R}_V$ shows a maximum for grAl sheet resistances in the range of $10-20$~$\Omega$, where we would expect the detector performance to be optimal, we find it remarkable that $\mathfrak{R}_V$ does not rapidly degrade at high resistivities, several orders of magnitude greater than the ones currently used in MKID technology. This slow decrease of $\mathfrak{R}_V$ with increasing kinetic inductance opens the way for the study, and possible use, of MKIDs with very large kinetic inductance ($\alpha \approx 1$).

\subsection{Measurement of the NEP} \label{SecNEPmeasurement}

%shining the resonators
The opening of the optical cryostat is coupled to a matte, high density polyethylene disk, which is cooled down by a pulse-tube cryocooler and used as a black body source. A layer of Eccosorb{\texttrademark} sponge can be manually interposed between the source and the cryostat, acting as a room temperature black body. This procedure allows to switch between a $100$ K and a $300$ K source. By accounting for this change in temperature, and the optical coupling between the source and the cryostat, we obtain the shift in radiant power on the sample $\delta P_\text{in}$ \cite{catalano2015bi}, in the range of $0.3$~pW for the H2 resonator design, and $0.1$~pW for the H3 design. For each of our samples we estimate the power coupled into the MKID, $P_\text{abs} = \delta P _\text{in} \cdot \mathcal{A}$, by calculating the film absorptance $\mathcal{A}$, using analytical formul{\ae } corroborated by finite element simulations (cf. Appendix~\ref{hilbertapp}).

Following Eq.~\eqref{respeq}, we obtain the MKID responsivity by measuring the resonant frequency shift, $\delta f_0$, when the illumination source changes from cryogenic black body to room temperature. The results for all samples are plotted in purple in Fig.~\ref{fig:detector_params}a. The brown curve in Fig.~\ref{fig:detector_params}a shows the measured shift in QP density for absorbed power, $\delta x_\text{qp} / P_\text{abs} $, obtained by using the second and third parts of Eq.~\eqref{respeq}. The QP signal is remarkably constant for all resistivities, with an average value $\langle \delta x_\text{qp} / P_\text{abs} \rangle \approx (1.2 \pm 0.4) \cdot 10^{-4}$~pW$^{-1}$, indicating that the measured fluctuations in $\mathfrak{R}$ are simply due to different fundamental mode frequencies, $f_0$, of MKIDs in different samples.

We calculate the NSD by recording $\delta f_0 (t)$ in the absence of incoming radiation, when the opening of the optical cryostat is covered. Since perfect optical sealing can not be achieved in the optical cryostat, we repeat the measurement in the dark cryostat to obtain the intrinsic noise figure of the MKIDs. The comparison between the two measured NSDs is shown in Fig.~\ref{fig:detector_params}b. Note that the dependence of the NSD versus resistivity shows a minimum for measurement performed in both cryostats, and the values measured in the dark are a factor of two lower. This non-monotonic dependence versus normal state resistivity of the film is correlated with the superconducting gap value reported in Fig.~\ref{fig:detector_params}d, which suggests quasiparticle generation-recombination as the dominant source of noise \cite{sergeev1996photoresponse,sergeev2002ultrasensitive,de2011number,de2012generation}. This would also explain the lower NSD measured in the dark cryostat, where the superior shielding and filtering results in a lower density of non-thermal QPs.

We compute the NEP as the ratio of the NSD measured in the optical cryostat and the responsivity $\mathfrak{R}$, as per Eq.~\eqref{nepeq}, and we plot the results in Fig.~\ref{fig:detector_params}c. The  dependence of the NEP vs. grAl film resistivity shows a minimum at $\rho \approx 200$~$\upmu \Omega \cdot$~cm, corresponding to a sheet resistance $R_n \approx 100$~$\Omega$, which is an order of magnitude larger than the typical values used in the MKID community \cite{catalano2015bi,janssen2014performance,bueno2017full,de2012generation,baselmans2016performance,flanigan2016photon}.

We plot the fitted values of the grAl superconducting gap $\Delta$ in Fig.~\ref{fig:detector_params}d in order to highlight its anticorrelation with the NEP. As discussed in Section~\ref{sec2PhononTrapping}, under the assumption of dominating QP generation-recombination noise, we expect a minimum in the NEP when the Al ground plane is most efficient in phonon trapping (i.e. when the grAl gap is maximum). 

Within this model, we expect an anticorrelation between NEP and $\Delta$, as given by Eq.~\eqref{neppr}. In Fig.~\ref{FractionalChangeNEP} we plot the measured fractional change in NEP vs. the fractional change in $\Delta$, as well as the expected dependence according to equation Eq.~\eqref{neppr}. For the theoretical line we used our measured averaged values ${2 \overline{\Delta_G} /h } \approx 160$~GHz and ${2\Delta_A/h} \approx 100$~GHz \cite{maleeva2018circuit}. Except for sample F, we do observe an anticorrelation  between the change in NEP and the height of the superconducting grAl gap, suggesting that phonon trapping in the Al ground plane plays an important role. The fact that sample F - the one with the highest resistivity - deviates the most from the model Eq. \eqref{neppr} could point to an additional complexity in the quasiparticles dynamics that is currently unaccounted for. Namely, the quasiparticle lifetime can increase from hundreds of microseconds in pure Al films \cite{swenson2010high} up to seconds in highly resistive grAl films \cite{grunhaupt2018dynamics}.

\section{Conclusions} \label{sec:conclusions}

We used granular aluminum (grAl) as a novel thin film material to fabricate MKIDs with resistivities ranging from $40$ to $1600\;\upmu\Omega\cdot\text{cm}$, corresponding to kinetic inductances up to orders of magnitude higher than those found in current MKID technology. To minimize the NEP, we found an interplay between kinetic inductance fraction $\alpha$ and the non-linearity limiting the maximum number of readout photons before bifurcation to $n_\text{max}$, resulting in an optimal $\alpha = 3/4$. This  value does not depend on resonator geometry, which can be optimized for maximum absorptance. In order to quantify the outcome of this interplay, we introduced the concept of voltage responsivity $\mathfrak{R}_V = \alpha \sqrt{n_\text{max}}$.  For $\alpha > 3/4$ we expect an increase of the NEP due the slow decrease of $\mathfrak{R}_V$; however, experimentally, we found the NEP to be minimum at $\alpha \approx 0.9$. This is due to the pronounced minimum of the NSD at $\alpha \approx 0.9$, which coincides with the region of maximum grAl superconducting gap as a function of resistivity. We explain the anticorrelation between NSD and grAl gap using a phonon trapping model in the surrounding Al ground plane. The measured NEP values for grAl MKIDs, scaled for maximum absorptance to allow for a fair comparison between different film resistivities, are in the range of $25$~aW$/\sqrt{\text{Hz}}$, and are comparable to state of the art \cite{ de2012generation,calvo2016nika2,flanigan2016photon,baselmans2016performance,bueno2017full}. 

Guided by these results, future research should focus on increasing the grAl superconducting gap, e.g. using a cold deposition method as found in Ref. \cite{cohen1968superconductivity}), employing a thicker and lower gapped  (e.g. titanium) ground plane, and engineering the meandered inductor geometry in order to maximize the optical impedance matching between the resonator film and the photon collecting medium. We believe that the flexibility and low losses of grAl constitute an advantage for future ultrasensitive MKID applications. 

\section*{Acknowledgements}

We are grateful to A. Karpov for insightful discussions, and to L. Radtke and A. Lukashenko for technical support. Facilities use was supported by the KIT Nanostructure Service Laboratory (NSL). Funding was provided by the Alexander von Humboldt foundation in the framework of a Sofja Kovalevskaja award endowed by the German Federal Ministry of Education and Research, and by the Initiative and Networking Fund of the Helmholtz Association, within the Helmholtz Future Project Scalable solid state quantum computing. This work was partially supported by the Ministry of Education and Science of the Russian Federation in the framework of the Program to Increase Competitiveness of the NUST MISIS, contracts no. K2-2016-063 and K2-2017-081.

\section*{APPENDICES}
\appendix

\counterwithin{figure}{section}

\section{Geometry choice and impedance matching} \label{hilbertapp}

\begin{figure}
\centering
\def\svgwidth{0.47\textwidth}  
\hspace{-0.25cm}
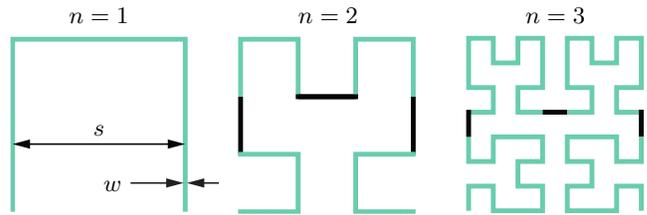
\caption{First three iterations of a Hilbert fractal, i.e. Hilbert curve of the first three degrees. In green we higlight the recurrence of $(n-1)$-th structures in the $n$-th Hilbert fractal iteration, connected by black lines.} \label{hilb}
\vspace{1cm}
\end{figure}

We give a brief overview of the choice of resonator geometry, following in the footsteps of the much more detailed treatment found in \cite{rosch2014development}. The detectors used in this work are back-illuminated lumped element resonators, composed by a meandered inductor and interdigitated capacitor, and patterned on sapphire. They offer a practical advantage over distributed resonators since the lumped element capacitance can be swept by changing the length of the capacitor fingers, with no effect on the inductance. This is needed to obtain a fine comb of resonant frequency dips, limited only by the loaded quality factor of the resonators, which is an important consideration towards densely packed kinetic inductance arrays. The meandered inductor is shaped as a Hilbert curve \cite{hilbert1891ueber}, shown in Fig.~\ref{hilb}. 

The $n$-th order fractal can be ideally decomposed into $4^n-1$ zeroth order structures, which are simple stripes of width $w$ and length $s+w \approx s$, oriented horizontally and vertically with approximately equal distribution. The filling factor is defined as $\text{ff}=w/s$. This geometry renders the detector sensitive to two polarizations at once. Increasing the degree $n$ increases the inductance, and decreases the impedance. The two geometries we used in this work are second and third degree Hilbert curves (H2 and H3), since $n=2$ and $n=3$ offer a good compromise between resonant frequencies that are within the bandwidth of the readout electronics and impedances that are easily matched to the substrate. We estimate the resonator sheet resistance per square needed to assure impedance matching as
\begin{equation}
R_{n,\text{match}} \cdot \frac{s}{w} = \frac{Z_0}{\sqrt{\epsilon_\text{sapphire}}} \iff R_{n,\text{match}} \approx \text{ff} \cdot 100 \; \Omega,
\end{equation}
where $Z_0 \approx 377 \; \Omega$ is the impedance of vacuum and $\epsilon_\text{sapphire}$ is the relative dielectric permittivity of the sapphire substrate. We compare this formula to finite element simulations, showing an accuracy within $5\%$, and use it to estimate the detector absorptance as 
\begin{equation}
\mathcal{A} (R_n) = \frac{|R_n - R_{n,\text{match}}|}{R_n + R_{n,\text{match}}},
\end{equation}
which we then use to calculate $P_\text{abs} = \delta P_\text{in} \cdot \mathcal{A}$. 
The fabricated MKIDs have $\ell = 2.5$ mm, $w = 2\; \upmu$m (H3) and $\ell = 1.2$ mm, $w=12\; \upmu$m (H2). Notice that H2 resonators have about three times larger surface.

\section{Noise spectral density}\label{NSDapp}
In MKIDs, the detector signal is the resonant frequency shift $\delta f_0$ caused by millimeter wave illumination. As a consequence, any oscillation of the resonant frequency observed in the absence of incoming millimeter wave photons constitutes noise. The noise is calculated by observing the resonant frequency oscillations over time $\delta f_0(t)$ under zero illumination condition and computing the noise spectral density (NSD) as 
\begin{equation} \label{nsd}
\text{NSD}(f) = \sqrt{    \frac{ | \mathcal{F} \{ \delta f_0 (t) \}  |^2}{\text{BW}} } = \frac{  |\widetilde{\delta f_0} (f)|}{ \sqrt{ \text{BW}}},
\end{equation}
where BW is the output bandwidth defined as $\text{BW} = 1/2\cdot t_ \text{exposure}$. We recall that the change in resonant frequency is caused by a change in kinetic inductance,
\begin{equation} \label{shiftapp}
\delta f_0  = - f_0 \frac{\alpha}{2} \frac{\delta L_\text{kin}} { L_\text{kin}}.
\end{equation}
The kinetic inductance scales with the inverse of the Cooper pair density $n_S$, allowing one to write
\begin{equation} \label{kintoqp}
-\frac{\delta L_\text{kin}} { L_\text{kin}} = \frac{\delta n_S}{n_S} = 2\delta x_\text{qp},
\end{equation}
where $x_\text{qp}$ is the normalized grAl quasiparticle density. Under the assumption of quasiparticles dominating over other sources of noise, we can then link the fluctuations of the resonant frequency over time with fluctuations of the quasiparticle density 
\begin{equation}
\delta f_0 (t) = \alpha f_0 \delta x_\text{qp}(t).
\end{equation}
The resonant frequency fluctuates by up to roughly $1$ kHz, thus for $\alpha \approx 1$ and $f_0$ of the order of GHz, we obtain $\delta x_\text{qp}\approx~10^{-6}$, comparable to the background values recorded in \cite{grunhaupt2018dynamics}.

\section{Phonon trapping model}\label{trappingmodel}

As a starting point, we take the equation
\be\label{NEPstart}
\mathrm{NEP} = 2\Delta_G \sqrt{\frac{N_G}{\tau_G}}
\ee
relating the NEP to the gap $\Delta_G$, the quasiparticle number $N_G$, and the quasiparticle lifetime $\tau_G$. From now on we use subscript $G$ for quantities in granular aluminum, $A$ for regular aluminum, and $P$ for phonons.

The system has in principle several parts. For example, there are phonons in the substrate, in $A$, and in $G$. We assume that all these phonons quickly reach a steady state under the experimental conditions, so we can collectively describe them with a single phonon distribution. Moreover, since temperature $T$ is small ($T \ll \Delta_G,\, \Delta_A,\,\Delta_G - \Delta_A$), we neglect any effect of thermal phonons and we focus on ``hot'' phonons, with energy above the highest $G$ gap ($E_P > \Delta_G^\mathrm{max}$), so they can create quasiparticles in $G$ by breaking Cooper pairs at a rate $b_G$ independent of $\Delta_G$. We indicate with $N_P$ the number of hot phonons. In addition to quasiparticles in $G$ with number $N_G$, we also have quasiparticles in $A$. We are interested in ``hot'' quasiparticles, generated by phonons; they can recombine by emitting a hot phonon again, or scatter by emitting lower-energy phonons that cannot break pairs in $G$.
We model the dynamics of hot phonons and quasiparticles in a phenomenological way, with rate equations of the Rothwarf-Taylor type. For quasiparticles in $G$, the relevant processes are generation from pair breaking by hot phonons (rate $b_G$) and recombination (rate $r_G$). Similarly, for quasiparticles in $A$ we have generation by pair breaking ($b_A)$ and recombination ($r_A)$, but also scattering to lower energies (rate $s_A$). For the phonons, we assume some generation mechanism with rate $g_P$, in addition to generation/recombination in both $G$ and $A$. The rate equations are then:
\bea
\dot{N}_G & = & -2r_G N_G^2 + 2 b_G N_P \label{NG_eq} \\
\dot{N}_A & = & -2r_A N_A^2 + 2 b_A N_P - s_A N_A\\
\dot{N}_P & = & g_P - b_A N_P + r_A N_A^2 - b_G N_P + r_G N_G^2 \label{NP_eq}
\eea
We consider now the steady-state solution. The first equation simply gives
\be\label{NG_sol}
N_G = \sqrt{b_G N_P/r_G}
\ee
and the last two terms in the last equation cancel out. Then we are left with the system
\bea
0 & = & -2r_A N_A^2 + 2 b_A N_P - s_A N_A\\h
0 & = & g_P - b_A N_P + r_A N_A^2
\eea
We can solve the first equation for $N_A$ in terms of $N_P$, substitute into the second equation, and find
\be
N_P = \frac{g_P}{b_A} + \frac{4r_A g_P^2}{b_A s_A^2}
\ee
For ``weak'' scattering, $s_A \ll 2\sqrt{r_A g_P}$, we then have
\be\label{NP_ws}
N_P \approx \frac{4r_A g_P^2}{b_A s_A^2}
\ee
Note that \eref{NP_ws} diverges as the scattering rate decreases. This unphysical result is due to the fact that we have neglected other mechanisms that can decrease the hot phonon number, for example escape from the substrate into the sample holder, or phonon scattering that cools them below the gap of $G$. In a relaxation-time approach, such a contribution would add a term $-e_P N_P$ to the right hand side of \eref{NP_eq} in order to account for escaped phonons. Then we can show that \eref{NP_ws} remains valid so long as $4e_P/b_A \ll s_A^2/4r_A g_P$.

Having $N_P$, we can calculate $N_G$ using \eref{NG_sol}. As for the quasiparticle lifetime, linearizing \eref{NG_eq} around the steady state, one can see that $1/\tau_G = 4 r_G N_G$. Therefore we get from \eref{NEPstart}
\be\label{NEPDN}
\mathrm{NEP} \approx 4\Delta_G \sqrt{b_G N_P} = 8g_P \sqrt{\frac{b_Gr_A}{b_A}}\, \frac{\Delta_G}{s_A} 
\ee
The quasiparticle scattering rate $s_A$ depends on the quasiparticle energy $\epsilon$ above the gap $\Delta_A$. At low temperature, we approximate the scattering rate due to electron-phonon interaction by its zero-temperature expression, which according to \cite{kaplanqp} can be written in the form:
\begin{widetext}
\be
s_A = \frac{1}{\tau_{0A} \Delta_A^3} \int_0^\epsilon d\omega \, \omega^2 \frac{\Delta_A +\epsilon - \omega}{\sqrt{(\Delta_A+\epsilon -\omega)^2-\Delta_A^2}} \left[1-\frac{\Delta_A^2}{(\Delta_A+\epsilon)(\Delta_A+\epsilon - \omega)}\right]
\ee
\end{widetext}
where the prefactor $1/\tau_{0A}$ accounts for the strength of the electron-phonon interaction.
For $\epsilon \lesssim \Delta_A$ we then find $s_A \propto (\epsilon/\Delta_A)^{7/2}/(1+\epsilon/\Delta_A)$, while for $\epsilon \gg \Delta_A$ we have $s_A \propto (\epsilon/\Delta_A)^3$. Here $\epsilon \approx \Delta_G-\Delta_A$ is of order $\Delta_A$, so the first expression applies. 
Since we are interested in the dependence of NEP on $\Delta_G$, dropping prefactors we arrive at
\be
\mathrm{NEP} \propto \Delta_G^2 \left(\frac{\Delta_A}{\Delta_G-\Delta_A}\right)^{7/2}
\ee
Therefore the relative change in NEP due to change in $\Delta_G$ is
\be\label{dNEPdDG}
\frac{\delta\mathrm{NEP}}{\mathrm{NEP}} = -\left(\frac{3\Delta_G+4\Delta_A}{2(\Delta_G-\Delta_A)}\right) \frac{\delta\Delta_G}{\Delta_G}
\ee
Average values of NEP and $\Delta_G/h$ are about $70\; \text{aW}/\sqrt{\text{Hz}}$  and $80$ GHz. We use these average values as reference points to calculate $\delta$NEP and $\delta\Delta_G$. We use the average $\Delta_G$ and $\Delta_A/h \approx 50\,$GHz for a 50\,nm thick Al film in \eref{dNEPdDG} to calculate the slope to be about $-7.3$. We estimate the uncertainty for the averaged values to be roughly $10\%$, which propagates as an uncertainty in the value of the slope.

\section{Kinetic inductance fraction} \label{alphaapp}

The kinetic inductance fraction $\alpha$ is defined as $\alpha = L_\text{kin}/(L_\text{kin}+L_\text{geom})$. The kinetic inductance of a superconducting film can be expressed as \cite{rotzinger2016aluminium,tinkham1996introduction,annunziata2010tunable}
\begin{equation}\label{lkin}
L_\text{kin} = \frac{0.18 \hbar}{k_B} \frac{\ell}{w} \frac{R_n}{T_c},
\end{equation}
where $\ell$ is the length, $w$ is the width, $R_n$ is the normal state sheet resistance per square and $T_c$ is the critical temperature. In the case of the loop-free meandered inductors employed in our resonators, the geometric inductance is only given by self inductance. While an exact closed formula for the self inductance of a rectangular bar can be computed (Ref.~\cite{piatek2012exact}), it is lengthy and cumbersome. A much more compact formula can be found in Ref.~\cite{terman1943radio}, which in the thin ribbon limit ($\ell \gg w \gg t$) reads 
\begin{equation} \label{lgeom}
L_\text{geom} \approx 2\cdot 10^{-7} \ell \ln \left( \frac{2\ell}{w}  \right).
\end{equation}
We tested this formula against the exact one for all combinations of $\ell$, $w$ and $t$ listed in Table \ref{testgeoms}. 
\begin{table} [h!]
\begin{center}
  \caption{Geometric parameters with which Eq. \ref{lgeom} was tested against the exact solution found in Ref.~\cite{piatek2012exact}.}\label{testgeoms}
  \vspace{0.3cm}
\begin{tabular}{|c|c|c|c|}
    \cline{2-4}
    \multicolumn{1}{c|}{} & length [$\upmu$m] & width [$\upmu$m] & thickness [nm] \\ \hline
    span &   $200-2000 \;  $ & $2-20 \;   {}$ & $10-50 \;   {}$ \\ \hline
    step &  $100 \;  {}$ & $2 \; {}$ & $ 5 \;  {}$ \\
    \hline
  \end{tabular}
  \end{center}
\end{table}
Errors were always below $10\%$. To estimate the kinetic inductance fraction $\alpha$ of our resonators starting from their geometry, we combine Eqs.~\eqref{lkin} and \eqref{lgeom}.

In Table \ref{litvals} we summarize a brief literature survey of previously measured kinetic inductance fractions for MKIDs made of various thin film materials. These values are used in Fig.~\ref{fig:alphanvoltresp} in the main text.  

\begin{table}[!h]
\caption{Summary of surveyed MKID parameters used in Fig.~\ref{fig:alphanvoltresp} in the main text, including both distributed and lumped element (LE) resonators.}\label{litvals}
\begin{center}
{\small{
\begin{tabular}{ | l | l | l| l | l | r| }
  \hline			
  Material &   LE & $R_n \; [\Omega]$ & $T_c \; [\text{K}]$ & $\alpha \; [\%]$ & Ref.   \\
    \hline
      \hline
  Al &   Yes & $0.66$ & $1.2$ &$6$ &\cite{cardani2015energy}  \\
    \hline
    Al &   No & $0.13$, $0.66$, $1.3$ & $1.2$ &$7, 45, 63$ &\cite{gao2008physics}  \\
      \hline
          Al &   No & $0.13$ & $1.2$ &$28 $ &\cite{gao2006experimental}  \\
          \hline       
           Al &   No & $0.13$ & $1.2$ &$7$ &\cite{mazin2006position}  \\
             \hline
            Al &   Yes & $4$ & $1.46$ &$40$ &\cite{mccarrick2014horn}  \\
             \hline
            Al &   Yes & $4$ & $1.2$ &$50$ &\cite{mauskopf2014photon}  \\
              \hline
             NbTiN &   No & $8.75$ & $14$ &$9$ &\cite{janssen2013high}  \\
               \hline
 NbTiN &   No & $5.6$ & $14$ &$35$ &\cite{yates2011photon}  \\
 \hline
  TiN &   Yes & $25$ & $4.1$ &$74$ &\cite{leduc2010titanium}  \\
   \hline
  TiN &   Yes & $45$ & $2$ &$\approx 100$ &\cite{swenson2013operation}  \\
   \hline
  WSi$_2$, W$_3$Si$_5$ &   No & $45,5.6$ & $1.8, 4$ &$92, 42$ &\cite{quaranta2013tungsten}  \\
    \hline
    WSi$_2$, W$_3$Si$_5$ &   Both & $7, 13, 44, 4.5$ & $1.8, 4$ &$43, 75, 96, 77$ &\cite{cecil2012tungsten}  \\
  \hline  
\end{tabular}
}}
\end{center}

\end{table}

\section{Maximum number of readout photons in grAl 1D resonators} \label{nmaxapp}
We give a brief recapitulation of the model proposed in Ref.~\cite{maleeva2018circuit}.
Granular aluminum (grAl) is a composite material made of aluminum grains in a non-stoichiometric aluminum oxide matrix. The structure of grains, separated by thin insulating barriers, is modeled as a network of Josephson junctions (JJs) \cite{hutter2011josephson,bourassa2012josephson,tancredi2013bifurcation,weissl2015kerr}. A stripline grAl resonator, or a lumped element grAl resonator with a thin enough meandered inductor, can be considered as a one-dimensional chain of effective JJs, which allows to derive the resonator self-Kerr \cite{walls2007quantum} non-linearity
\begin{equation} \label{app_selfkerr}
K_{11} = \mathcal{C} \pi e a \frac{\omega_0^2}{j_c V},
\end{equation}
where $a$ is the characteristic size of an aluminum grain, $V = \ell \cdot w \cdot t$ is the volume of the resonator, $j_c$ is the critical current density, $e$ is the electron charge, $\omega_0 = 2\pi f_0$ is the resonant frequency, and $\mathcal{C}$ is a geometric parameter of order unity, which in our case is $\mathcal{C}=3/16$.
The maximum number of photons in the resonator at bifurcation is \cite{eichler2014controlling}
\begin{equation} \label{app_nmax}
n_\text{max} = \frac{\kappa}{\sqrt{3}K_{11}},
\end{equation}
where $\kappa = f_0/Q_\text{tot}$ is the instantaneous bandwidth of the resonator. 
The total kinetic inductance of the JJ array is $L_\text{kin} = L_J \cdot \ell / a$, where $L_J$ is the inductance of a single effective JJ, allowing us to write
\begin{equation} \label{app_jc}
j_c = \frac{\ell \hbar}{2eaL_\text{kin}wt}.
\end{equation}
Under the assumption of strongly overcoupled resonators ($Q_\text{tot} \approx Q_c$), and kinetic inductance dominating over the geometric inductance ($1/f_0 \approx 2\pi \sqrt{L_\text{kin}C}$), we can use Eqs.~\eqref{app_selfkerr}~to~\eqref{app_jc} to write
\begin{equation}
n_\text{max} = \frac{4 \ell^2 \hbar \sqrt{C}}{ 3\sqrt{3}Q_c (\pi e a)^2 \sqrt{L_\text{kin}}}.
\end{equation}

\section{Sample holder design} \label{sampleholderapp}

\begin{figure}
\centering
\def\svgwidth{0.5\textwidth}  
\hspace{-0.5cm}
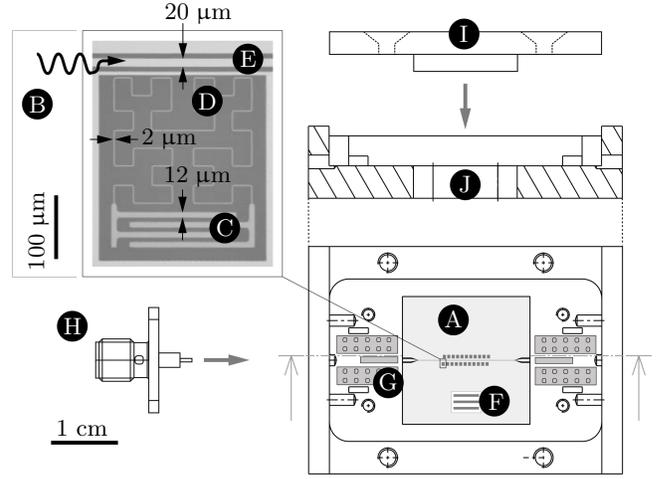
\caption{Technical drawing of the aluminum sample holder. The inset shows an optical image of a single H3 resonator. } \label{comp}
\end{figure}

We show a technical drawing of the aluminum sample holder in Fig. \ref{comp}.
The inset zooms in on the sapphire chip (A) to show one of the 22 resonators (B), where we highlight the interdigitated capacitor (C), meandered third degree Hilbert curve inductor (D) and CPW feedline (E). Note that H2 resonator have a meander width of $12\;\upmu$m. The chip also hosts test stripes (F) used for room temperature DC measurements of the sheet resistance (more stripes are present on the wafer prior to dicing). The feedline is wire bonded to the printed circuit boards (G) that couple to co-axial connectors (H). The sample holder is closed with a solid aluminum lid (I), with an aperture on the backside (J), allowing for mm-wave illumination.

\section{Photon number calibration in the dark cryostat} \label{nbarapp}
%\vspace{-1 mm}
We give a brief description of the dark cryostat experimental setup, and of the method used to estimate the number of photons circulating in a resonator.
\begin{figure}[!t]
%\vspace{2 mm}
\centering
\def\svgwidth{0.9\columnwidth}  
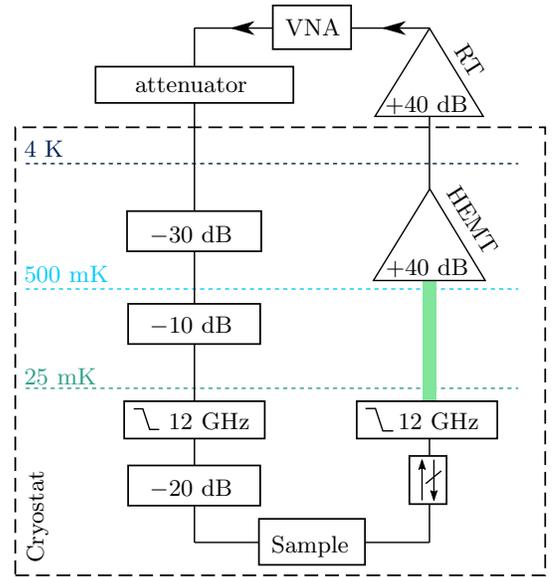
\caption{Schematic diagram of the measurement setup in the dark cryostat.} 
\label{photcalib}
\end{figure}
%\vspace{10 mm}
%
As schematically shown in Fig.~\ref{photcalib}, the sample under test is thermally anchored to the dilution stage of a cryostat, and we perform RF transmission measurements. We use a Vector Network Analyzer (VNA) to generate the input tone and to analyze the output. We add room temperature and cryogenic attenuators to the input line, with a typical total attenuation in the range of $-90$~dB. A number of fixed attenuators and a low pass filter are used at different temperature stages of the cryostat to prevent RF heating and to thermalize the input RF field. Once the RF signal is transmitted through the sample, an isolator is used to prevent back-propagating noise from the amplifiers. The signal is filtered and travels through superconducting cables (green) before being amplified by a high electron mobility transistor (HEMT) amplifier at $4$~K, and a room temperature amplifier (RT).

We estimate the drive power at the sample holder input, $P_\text{cold}$, simply by adding all known attenuation sources on the input line. For a strongly overcoupled resonator ($Q_\text{tot}  \approx Q_c$), we estimate the average number of photons using the following expression \cite{weber2011single}:
\begin{equation}
\bar{n} = P_\text{cold} \frac{2Q_c}{\hbar \omega_0^2}.
\end{equation}
For example, our typical values are $Q_c = 10^5$ and $f_0 = 5$ GHz, resulting in a one photon regime for $P_\text{cold} \approx -150$ dBm. The attenuation figures of the employed components and of the microwave lines, as well as the amplifiers gain, can only be coarsely estimated. For this reason, we expect our photon number estimation to be precise within an order of magnitude. Furthermore, as reported in the main text, this method appears to systematically underestimate the number of photons circulating in the resonators by a factor two, which might be due to overestimation of the input line attenuation.

\bibliography{NEPrefe}

\end{document}